\documentclass[11pt]{article}
\usepackage[utf8]{inputenc}
\usepackage{graphicx}
\usepackage{amssymb}
\usepackage{amsmath}
\usepackage{amsfonts}
\usepackage{wasysym}
\usepackage{textcomp}
\usepackage{ dsfont }
\usepackage{circuitikz}
\usepackage{ mathrsfs }
\usepackage{cancel}
\usepackage{circuitikz}
\usepackage{csquotes}
\usepackage[english]{babel}	
\usepackage{filecontents}
\usepackage{simpler-wick}

\usepackage{colortbl}
\usepackage[
sorting=none,
style=ieee, 
citestyle=numeric-comp,
backend=biber
]{biblatex}
\bibliography{bib2Dqg}

\bibliography{refs}

\usepackage{xcolor}
\usepackage{color} % подключить пакет color
\definecolor{BlueGreen}{RGB}{49,152,255}
\definecolor{Violet}{RGB}{120,80,120}
\definecolor{Blue}{RGB}{0,0,255}
\definecolor{ForestGreen}{RGB}{34, 139, 34}
\definecolor{Yellow}{RGB}{0,255,51}
\definecolor{ElectricGreen}{RGB}{0, 255, 0}
\definecolor{Indigo}{RGB}{75, 0, 130}
\definecolor{MediumPersianBlue}{RGB}{0, 103, 165}
\definecolor{PrussianBlue}{RGB}{0, 49, 83}
\definecolor{Fandango}{RGB}{181, 51, 137}
\definecolor{MediumPersianBlue}{RGB}{0, 103, 165}
\usepackage[unicode, colorlinks, urlcolor=Blue, linkcolor=MediumPersianBlue, citecolor=ForestGreen]{hyperref}

\usepackage[left=2cm,right=2cm,top=2cm,bottom=2cm,bindingoffset=0cm]{geometry}

\usepackage[T1, T2A]{fontenc}			% Пакет выбора кодировки и шрифтов
\usepackage[utf8]{inputenc} 			% любая желаемая кодировка

\usepackage{authblk}

\title{\bf On generalization of 2d gravity with the simplest non-invariant states}
\author[1,2]{Damir~Sadekov\thanks{\href{mailto:sadekov.de@phystech.edu}{sadekov.di@phystech.edu}}}
\affil[1]{\textcolor{black}{Institutskii per, 9, Moscow Institute of Physics and Technology, 141700, Dolgoprudny, Russia}}
\affil[2]{\textcolor{black}{B. Cheremushkinskaya, 25, Institute for Theoretical and Experimental Physics, 117218, Moscow, Russia}}
\date{\today}

\begin{document}

\maketitle

\begin{abstract}

Two-dimensional quantum gravity has led to numerous curious results since it was developed in the 1980s. Following the method of the original works, we derive the effective action for the simplest modifications of the theory, when the matter is immersed into Poincaré non-invariant states. The latter ones, as it has been shown recently, may have a considerable response in the theories with gravity, and the 2d Universe is a basic example to start with the deep investigation of these phenomena. We consider the cases of the insertion of the primary state of CFT and the thermal distribution of conformal matter.

\end{abstract}

\newpage

\tableofcontents

\newpage

\section{Introduction}\label{Intro}
In the 1980s A.M. Polyakov, V.G. Knizhnik and A.B. Zamolodchikov \cite{Polyakov2d, KPZ} exactly solved the two-dimensional quantum gravity coupled to a conformal matter with values of its central charge beyond the range $1<d<25$. However, the continuous approach in the phase of strong gravity is still an open question. The starting point was the formulation by A. Polyakov of Liouville theory of one scalar field as the theory of 2d gravity, induced by some CFT, where the scalar field plays the role of Weyl parameter. Further development of Polyakov-Liouville theory led to plenty of curious results such as exact correlation functions for different conformal operators in such a theory \cite{Zamolodchikov:1995aa, Zamolodchikov2, Belavin}, applications to matrix models, etc. In those considerations, correlation functions were calculated over the Poincaré invariant state. However, it was noticed recently that different initial states lead to a variety of new and interesting phenomena \cite{Akhmedos1,Akhmedos2,Akhmedos3}. We are interested to go back to the roots of the problem and find a model example of 2d theory of gravity, induced by conformal matter immersed in some states, which are not Poincaré invariant. After all, there is nothing special in Poincaré invariant states, once we consider curved spaces and quantum gravity. Recall that Poincaré symmetry is just the isometry of flat space. We of course start with the simplest examples.

Let us briefly review the original approach to the problem in question \cite{Polyakov, Polyakov2d, BosonStr}. One considers the following functional integral:
\begin{equation}\label{F1}
	\int \frac{\mathscr{D}g_{\mu\nu}}{\text{Vol.Diff}}e^{-W_{\text{eff}}[g_{\mu\nu}]} = \int  \frac{\mathscr{D}g_{\mu\nu}\mathscr{D}\Phi}{\text{Vol.Diff}} e^{-S_{\text{CFT}}[\Phi, g_{\mu\nu}]},
\end{equation}
where $\Phi$ is some conformal matter interacting with $g_{\mu\nu}$. One way to proceed is to consider small perturbations around flat space in general form, such that
\begin{equation}\label{F2}
	\delta S_{\text{CFT}} = -\frac{1}{4\pi}\int d^2x \sqrt{|g|} T_{\mu\nu}h^{\mu\nu}, \;\; g_{\mu\nu} = \eta_{\mu\nu}+h_{\mu\nu},
\end{equation}
where $T_{\mu\nu}$ is the stress-enegry tensor and $\eta_{\mu\nu}$ is the flat metric. Here and below we use the light-cone notations and Euclidean signature, such that $\mu,\nu = +,-$ and $x_{\pm}=x_0\pm i x_1$, where $x_0, x_1$ are Cartesian coordinates. Then one takes the relation (\ref{F1}), expands in powers of $h_{\mu\nu}$ and calculates the effective action perturbatively. In the second order for the spherical topology one gets
\begin{equation}\label{F3}
	W_{\text{eff}}^{(2)} = \frac{d}{96\pi}\int d^2x d^2y R(x)\frac{1}{4\pi}\log\left(|x-y|^2\right)R(y),
\end{equation}
where $d$ stands for the central charge of the conformal theory under consideration and $R$ is the scalar curvature up to the first order:
\begin{equation}\label{F4}
	R \simeq 4\left( \partial_{+}^2h_{--}+\partial_{-}^2h_{++}-2\partial_{+}\partial_{-}h_{+-} \right).
\end{equation}
Note that $h_{+-}$ carries the dependence on Weyl parameter, while this dependence is absent at the classical level due to the tracelessness of the stress-energy tensor in the flat background. The crucial point in the calculation under consideration is the restoration of the Weyl factor at the quantum level due to local counterterms originated in Pauli-Villars regulators, which should be added to the theory. The full effective action can be written down on the symmetry grounds \cite{Polyakov} (see also section \ref{Symm}):
\begin{equation}\label{F5}
W_{\text{eff}} = S_{P} = \frac{d}{96\pi} \int_M R\frac{1}{\Box}R,
\end{equation}
where $\frac{1}{\Box}$ solves the equation:
\begin{equation}\label{F6}
\Box_{x}\frac{1}{\Box}(x,y) \overset{\text{def}}{=} \frac{1}{\sqrt{|g|}}\partial_{\mu}\sqrt{|g|}g^{\mu\nu}\partial_{\nu}\frac{1}{\Box}(x,y) = \frac{1}{\sqrt{|g|}}\delta^{(2)}(x-y)
\end{equation}
and $M$ is some manifold with the topology of the sphere. Further consideration of two-dimensional gravity at the quantum level encounters the difficulty of non-linear integration measure. Despite being possible to partly overcome this problem in the conformal gauge $ds^2=e^{\varphi}dx^{+}dx^{-}$ via Distler and Kawai conjecture \cite{Distler:1988jt},  we will use a different approach. Namely, following A.M. Polyakov \cite{Polyakov2d, KPZ, LesHouches}, in such a problem it is convenient to use the light-cone gauge (see also \cite{Grund, Suzuki:1990wg}):
\begin{gather}\label{F7}
ds^2 = dx^+ dx^- + h_{++}(dx^{+})^2,
\\
||g_{\alpha\beta}|| =
\begin{pmatrix}
h_{++} & \frac{1}{2}\\
\frac{1}{2} & 0
\end{pmatrix}, \;\; \sqrt{|g|}=\frac{1}{2}. \nonumber
\end{gather}
In fact, compare the norms \cite{Polyakov, BosonStr}: 
\begin{equation*}
    ||\delta \varphi||^2 =  \int e^{\varphi} \left(\delta \varphi\right)^2 d^2x
\end{equation*}
and
\begin{equation*}
    ||\delta h_{++}||^2 =  \int \left(\delta h_{++}\right)^2 d^2x,
\end{equation*}
so the fluctuating variable $h_{++}$ has such a measure as that of the ordinary scalar field (while $\varphi$ does not) and the theory can be treated properly at quantum level \cite{LesHouches}. Then the effective action is defined as:
\begin{equation}\label{F8}
e^{-W_{\text{eff}}} \overset{\text{def}}{=} \left\langle e^{-\frac{1}{\pi}\int T_{--}h_{++}}\right\rangle_{\text{conf. matter}}.
\end{equation}
Note that in general the expression (\ref{F5}) is completely determined by the conformal algebra and is independent of the given CFT (up to an overall factor). The resulting effective action in the gauge (\ref{F7}) is:
\begin{equation}\label{F9}
	W_{\text{eff}} = \frac{k_0}{4\pi}\int d^2xd^2y \partial_{-}^2h_{++}(x)\frac{1}{\partial_{-}\left(\partial_{+}-h_{++}\partial_{-}\right)}\partial_{-}^2h_{++}(y),
\end{equation}
where $k_0=\frac{d}{6}$ classically and actually gets renormalized at the quantum level \cite{KPZ, LesHouches}. However, in the present paper we will neglect this effect of central charge renormalization formally assuming a quasi-classical limit $d\rightarrow +\infty$ and leave the accurate discussion of this phenomenon in the context of our modifications for further work. 

There are some interesting points to note about the action (\ref{F9}). First, it can be thought of as a gravitational analogue of WZNW action, which appears as an effective action in the theory with fermions coupled to gauge potential \cite{Wiegmann}. Second, it is possible to write down Ward identities \cite{Polyakov2d, Polyakov}. In fact, in order to find the variation of (\ref{F9}) we need to introduce the change of the variable $h_{++} = \frac{\partial_{+}f}{\partial_{-}f}$, which allows one to determine the exact Green function (see Appendix \ref{AppA}):
\begin{equation}\label{F10}
	\frac{1}{\Box}(x,y) = \frac{1}{4\pi}\log\big\{(x_{+} - y_{+})\left(f(x)-f(y)\right)\big\}.
\end{equation}
Now, combining the variation of these fields under ``anti-chiral'' gauge transformation
\begin{equation}\label{F11}
	\begin{cases}
		\delta f = \epsilon_{+}\partial_{-}f \\
		\delta h_{++} = (\partial_{+} - h_{++}\partial_{-})\epsilon_{+} + (\partial_{-}h_{++})\epsilon_{+}
	\end{cases}
\end{equation}
with the equations of motion for the action (\ref{F9}) (see \cite{Polyakov2d})
\begin{equation}\label{F12}
	-\frac{c}{24\pi}\partial_{-}^{3}h_{++} = 0, \;\; c=26-d,
\end{equation}
we write down the Ward identities:
\begin{gather}
\frac{c}{24\pi i} \partial_{-}^3 \Big\langle h_{++}(z)h_{++}(x_1)...h_{++}(x_N)\Big\rangle = 
\sum\limits_{k=1}^{N}\partial_{+}\delta(z-x_k)\Big\langle h_{++}(x_1)...\cancel{h_{++}}(x_k)...h_{++}(x_N) \Big\rangle + \nonumber
\\
+ \sum\limits_{k=1}^{N}\left[\delta(z-x_k)\frac{\partial}{\partial x_k^{-}} - \partial_{-}\delta(z-x_k)\right]\Big\langle h_{++}(x_1)...h_{++}(x_N)\Big\rangle.
\label{F13}
\end{gather}
\\
In this paper, we perform the same procedure, provided that the expectation value in the CFT is taken over the simplest initial states, which are not Poincare invariant.
\section{CFT with perturbed state}\label{PartitionFunction}
In our present discussion, we drop the functional integral over the metrics $g_{\mu\nu}$ and assume that expectation values in CFT are taken over some non-trivial density matrix $\hat{\rho}_{0}$. Namely, we consider the following partition function instead of (\ref{F1}):
\begin{equation}\label{F14}
	\mathcal{Z}[g_{\mu\nu}]=  \left\langle e^{-S_{\text{CFT}}[\Phi, g_{\mu\nu}]} \right\rangle_{\text{matter, }\hat{\rho}_0}.
\end{equation}
For instance, it may be built from in- and out- states, created by the insertions of conformal operators $\hat{\rho}_{0} = \sum\limits_{ij} A_{ij}|\hat{O}_i \rangle \langle \hat{O}_j|$, or we can take the thermal distribution of conformal matter etc. Thus, the leading-order effective action should be obtained from the expansion
\begin{equation}\label{F15}
e^{-W_{\text{eff}}} = \left\langle e^{-\frac{1}{\pi}\int T_{--} h_{++}}\right\rangle_{\hat{\rho}_{0}} =  1 - \frac{1}{\pi}\int h_{++}\langle T_{--}\rangle_{\hat{\rho}_{0}} +\frac{1}{2\pi^2}\int h_{++}\langle T_{--}T_{--}\rangle_{\hat{\rho}_{0}} h_{++} + \ldots\;,
\end{equation}
where averages are taken over free CFT with the insertion of $\hat{\rho}_0$ rather than Poincare-invariant vacuum as in (\ref{F8}). As we will see in the examples below, the sum of this series differs from the classical result (\ref{F5}). 

Finally, before we proceed with particular calculations, note that, in fact, the light-cone gauge (\ref{F7}) is well-defined only for Minkowski signature, such that $ds^2$ is a real number. Moreover, the component $T_{--}$ of the stress-energy tensor is a complex quantity in Euclidean signature, which depends on the complex variable $x_{-}$. However, as long as we consider the perturbative expansion (\ref{F15}), where each term is averaged over free CFT, the functional integral is defined correctly, because the conformal field theory is well-defined in flat space with Euclidean signature. Eventually, after the resummation of the series (\ref{F15}), one can restore the whole expression in terms of covariant quantities and find the answer in arbitrary gauge.
\subsection{The simplest case -- one primary state}\label{PrimaryState}
We start the treatment of (\ref{F15}) with the consideration of the simplest non-trivial case of pure initial spinless primary-state: $\rho_{0} = |\Delta\rangle \langle \Delta|$. We define this state and its conjugate in the standard way as energy eigenstates in Hamiltonian formalism in CFT, where time coordinate $\tau$ is defined as $z=e^{\tau-i\sigma}$ and
\begin{equation}\label{F16}
	|\Delta\rangle \overset{\text{def}}{=} \lim\limits_{z\rightarrow 0} \Phi_{\Delta}(z)|0\rangle, \;\;
	\langle \Delta | = |\Delta\rangle^{\dagger} = \lim\limits_{z\rightarrow 0} \langle 0 | \frac{1}{z^{2\Delta}} \Phi_{\Delta}\left( \frac{1}{z} \right).
\end{equation}
Let us stress that we consider $h_{++}$ here as a non-dynamical background field and calculate the integral over conformal fields on the sphere with two fixed punctures at $z=0$ and $\infty$, but we don't investigate correlation functions in $2d$ quantum gravity, where one should perform the dressing of operators by the metric to invert them into invariant objects \cite{Distler:1988jt}. Furthermore, with dynamical gravity for generic initial state one is ought to use Schwinger-Keldysh rather than Feynman diagrammatic technique and Minkowski rather than Euclidean signature.

In order to investigate the expansion (\ref{F15}) for such a state up to the second order in $h_{++}$ we find correlation functions of the stress-energy operator:
\begin{equation}\label{F17}
	\langle \Delta| T_{--}(x) |\Delta\rangle = \frac{\Delta}{(x_{-})^2},
\end{equation}
\begin{equation}\label{F18}
	\langle \Delta| T_{--}(x)T_{--}(y) |\Delta\rangle = \frac{d/2}{(x_{-}-y_{-})^4} + \frac{2\Delta}{x_{-}y_{-}(x_{-}-y_{-})^2} +\frac{\Delta^2}{(x_{-})^2(y_{-})^2}.
\end{equation}
Hence, in the first two orders the expansion looks as
\begin{gather}
	e^{-W_{\text{eff}}} = 1-\frac{\Delta}{\pi}\int d^2\boldsymbol{x}h_{++}(x)\frac{1}{(x_{-})^2}  + \frac{\Delta}{\pi^2}\int d^2\boldsymbol{x}d^2\boldsymbol{y}h_{++}(x) \frac{1}{x_{-}y_{-}(x_{-}-y_{-})^2} h_{++}(y) + \nonumber \\
	+\frac{1}{2}\left(\frac{\Delta}{\pi}\int d^2\boldsymbol{x}h_{++}(x)\frac{1}{(x_{-})^2}\right)^2 - S_{P}^{(2)} + \ldots\;,
\label{F19}
\end{gather}
where $S_{P}^{(2)}$ is the second order of the Polyakov's action and bold symbols stand for the pair of Cartesian coordinates $(x_0, x_1)$. From the other hand, one can expand the Green function in the presence of gravitational background to obtain (Appendix \ref{AppA}):
\begin{equation}\label{F20}
	\frac{1}{\Box}(x,y) = \frac{1}{4\pi}\log\left( (x_{+}-y_{+})(x_{-}-y_{-}) \right) + \frac{1}{4\pi^2}\int d^2\boldsymbol{z}\frac{1}{x_{-}-z_{-}}\frac{1}{z_{-}-y_{-}}h_{++}(z) + \ldots\;\;.
\end{equation}
Then, using this Green function, one can decompose the following expression
\begin{gather}
	e^{\Delta \int\limits_{M}\frac{1}{\Box}(x,0)R(x) - S_{P}} = 1 + \frac{\Delta}{\pi}\int d^2\boldsymbol{x} \partial_{-}^2 h_{++}\log\left( (x_{+}-y_{+})(x_{-}-y_{-}) \right) - S_{P}^{(2)} - \nonumber \\
	+ \frac{\Delta}{\pi^2}\int d^2\boldsymbol{x}d^2\boldsymbol{y}\frac{\partial_{-}^2 h_{++}(x)}{x_{-}-y_{-}}\frac{h_{++}(y)}{y_{-}} + \frac{1}{2}\left( \frac{1}{\pi}\int d^2\boldsymbol{x} \partial_{-}^2 h_{++}\log\left( (x_{+}-y_{+})(x_{-}-y_{-}) \right)\right)^2+\;\text{higher orders}.
\label{F21}
\end{gather}
After integrating by parts and symmetrization of the terms in (\ref{F21}) over $x$ and $y$ one can see that (\ref{F21}) coincides with (\ref{F19}). In the same way, one can show the coincidence in the third order in $h_{++}$. Thus, we propose the following result for the effective action
\begin{equation}\label{F22}
	W_{\text{eff}} = -\Delta \int d^2x \sqrt{|g|} \frac{1}{\Box}(x,0)R(x) + \frac{d}{96\pi}\int d^2x d^2y\sqrt{|g(x)|}\sqrt{|g(y)|} R(x)\frac{1}{\Box}(x,y)R(y).
\end{equation}
The action (\ref{F22}) describes Polyakov-Liouville gravity with an additional term, which is nothing but the Liouville field $\varphi(x) = -\int dy \frac{1}{\Box}(x,y)R(y)$, coupled to the delta-functional source at the origin with the charge $\Delta$. The symmetry argument of the next subsection makes those conclusions clear. 
\subsection{The symmetry argument}\label{Symm}
Since the theory is local and has more transparent geometric form in terms of the Liouville field it is instructive to consider the situation in conformal gauge. The locality is ensured by the fact that the free conformal field theory is classically Weyl-invariant, but the dependence on the Liouville field is restored due to UV-regularization, which is a short-distance effect. An interesting point about the effective theory with the non-trivial state under consideration is that the set of counterterms must be modified as well to assure general covariance and restoration of Liouville field dependence. 

Consider e.g. scalar conformal field theory and the primary state, created by the currents in holomorphic and anti-holomorphic sectors $J_{+} = i\partial_{+}\phi(x), \; J_{-} = i\partial_{-}\phi(x)$ according to the definition (\ref{F16}). The components of the stress-energy tensor in this case are:
\begin{equation}\label{F25}
	T_{++} = -\frac{1}{2}:\left(\partial_{+}\phi\right)^2:, \; T_{--} = -\frac{1}{2}:\left(\partial_{-}\phi\right)^2:, \; T_{+-} = T_{-+} = 0.
\end{equation}
Hence, using Wick theorem we can calculate one-point function for $T_{--}$ as
\begin{gather}
    \langle J_{+}J_{-}| T_{--}(x)|J_{+}J_{-}\rangle = \langle J_{+}|J_{+}\rangle\times2\times\frac{1}{2}\lim\limits_{z\rightarrow \infty} z^2 \langle \wick{ \partial_{-}\c\phi(z):\partial_{-}\c\phi(x) \partial_{-}\c \phi(x):\partial_{-} \c \phi(0) }\rangle = \nonumber \\
    =\lim\limits_{z\rightarrow \infty}\frac{z^2}{(z-x_{-})^2} \frac{1}{(x_{-})^2} = \frac{1}{(x_{-})^2}
\label{F26}
\end{gather}
and similarly for $T_{++}$. Then we obtain the effective action in the first order:
\begin{gather*}
    	W_{\text{eff}}^{(1)} = \frac{1}{\pi}\int d^2\boldsymbol{x} \left[h_{++}(x)\frac{1}{x_{-}^2} + h_{--}\frac{1}{x_{+}^2}\right] = -\frac{1}{\pi}\int d^2\boldsymbol{x} \log(x_{+}x_{-})\left[\partial_{-}^2h_{++}+\partial_{+}^2 h_{--}\right].
\end{gather*}
It is obvious now that we need to introduce additional terms even in the first order to guarantee the invariance under the gauge transformations $\delta h_{\mu\nu} = \nabla_{\mu}\epsilon_{\nu} + \nabla_{\nu}\epsilon_{\mu}$:
\begin{equation}\label{F27}
	W_{\text{eff}}^{(1)} = -\frac{1}{\pi}\int d^2\boldsymbol{x} \log(x_{+}x_{-})\left[\partial_{-}^2h_{++}+\partial_{+}^2 h_{--}-\boxed{2\partial_{+}\partial_{-} h_{+-}}\right].
\end{equation}
Despite the fact that the situation is much more complicated in higher orders we can use the property of conformal invariance of the theory. It implies that the effective action must be invariant under the analytical diffeomorphisms with the corresponding transformation of the Liouville field:
\begin{gather}
	ds^2 = e^{\varphi(z,\bar{z})}dzd\bar{z} \longrightarrow ds^2 = e^{\varphi(z,\bar{z})}\left| \frac{dz}{dw}\right|^2dwd\bar{w}, \nonumber
	\\
	\varphi(z,\bar{z}) \longrightarrow \varphi(w(z),\overline{w(z)}) + \log \left|\frac{dw}{dz}\right|^2.
\label{F23}
\end{gather}
However, we should modify this argument for the partition function (\ref{F14}), because the inserted density matrix $\rho_{0} = |\Delta\rangle \langle \Delta|$ relies on the chosen coordinate system as well: we insert the state $| \Delta \rangle$ at the origin and at infinity. Indeed, primary fields have an elementary transformation law $\Phi_{\Delta}(z) \rightarrow \left( \frac{dw}{dz} \right)^{\Delta}\Phi_{\Delta}(w(z))$. Hence, in order to ensure the invariance of the partition function with the insertions of the states (\ref{F16}) we should impose an additional requirement for allowed diffeomorphisms: they should preserve the origin $w(0) = 0$. This requirement is the manifestation of the broken Poincare symmetry of the problem in question. Therefore, in view of the transformation properties of primary fields $\Phi_{\Delta}(0) \rightarrow \left( \frac{dw}{dz} \right)^{\Delta}\Phi_{\Delta}(w(0))$ and the definition (\ref{F15}) of $W_{\text{eff}}$, i.e. in the form $e^{-W_{\text{eff}}} = \langle \Delta | \ldots |\Delta \rangle$, we conclude that the normalization factor of the partition function (\ref{F14}) with the inserted states (\ref{F16}) should be adjusted after the change (\ref{F23}) as follows:
\begin{equation*}
    e^{-W_{\text{eff}}[g_{\mu\nu}]} \longrightarrow \left|\frac{dw}{dz}\left( 0\right)\right|^{2\Delta} e^{-W_{\text{eff}}[\tilde{g}_{\mu\nu}]}.
\end{equation*}
Note that the transformation at $z=\infty$ is irrelevant, because it factors out in each term in (\ref{F15}) as in (\ref{F26}). In order to fit our requirements, one should write down the effective action in the following form:
\begin{equation}\label{F24}
	W_{\text{eff}} = \Delta \varphi(0) + A\int d^2x\left[\partial_{+}\varphi(x) \partial_{-}\varphi(x) + \mu^2e^{\varphi(x)}  \right],
\end{equation} 
where the coefficient $A$ can be determined from the perturbative expansion and the last term in the brackets in (\ref{F24}) stands for a possible cosmological term, which we will not discuss in this work. Hence, the effective action for the Liouville field is:
\begin{equation}\label{F28}
	W_{\text{eff}} = \Delta \varphi(0) - \frac{d}{48\pi}\int d^2x\frac{1}{2}\left(\partial_{\mu}\varphi\right)^2,
\end{equation}
in accordance with the result (\ref{F22}) of the previous subsection. 
\subsection{Including descendants}\label{Descendants}
So far we have discussed only primary states. However, in general one can include the whole spectrum of states of conformal matter in (\ref{F14})--(\ref{F15}) by considering the density matrix with some number of descendants of primary state $\Delta$:
\begin{equation}\label{F29}
	\hat{\rho} = \sum\limits_{\Delta^{\{k\}}, \Delta^{\{l\}}}A^{\Delta}_{\{k\}\{l\}}|\Delta^{\{k\}}\rangle \langle\Delta^{\{l\}}|, \;\; |\Delta^{\{k\}}\rangle \overset{\text{def}}{=} L_{-k_N}\ldots L_{-k_1}|\Delta\rangle,
\end{equation}
where the coefficients $A^{\Delta}_{\{k\}\{l\}}$ are normalized such that $\text{tr}\hat{\rho} = 1$. The calculation of an arbitrary matrix element of correlation functions of the stress-energy tensor over different states of the spectrum is rather difficult task, and there is no doubt that new terms in effective action will appear. Indeed, consider the simplest descendant state $|\Delta^{(1)}\rangle = L_{-1}|\Delta\rangle$ and see that the new type of summands arise already at the first loop:
\begin{equation}\label{F30}
	\frac{1}{2\Delta}\langle \Delta^{(1)}| T_{--}T_{--} |\Delta^{(1)}\rangle = \frac{d/2}{(x_{-}-y_{-})^4} + \frac{2\left(\Delta + 1\right)}{x_{-}y_{-}(x_{-}-y_{-})^2} +\frac{\left(\Delta + 1\right)^2}{(x_{-})^2(y_{-})^2} + 2\Delta\left[ \frac{1}{x_{-}y_{-}^3}+\frac{1}{x_{-}^3y_{-}} \right].
\end{equation}
Obviously, new types of summands are related to the fact that insertion of a state breaks another symmetries of partition function (\ref{F14}). Indeed, the general conformal operator $\hat{O}(x)$ transforms under an infinitesimal analytic change in the following way:
\begin{equation}\label{F31}
	\delta_{\epsilon}\hat{O}(x) = \sum\limits_{n=0}^{N} \hat{O}^{(n-1)}(x)\partial^{n}\epsilon(x) = \hat{O}^{-1}\epsilon + \hat{O}^{(0)}\partial \epsilon + \hat{O}^{(1)}\partial^{2}\epsilon + \ldots\;
\end{equation}
for some finite $N$ and other local operators $\hat{O}^{(n-1)}$. General descendant forbids more diffeomorphisms and hence provides us with more ingredients for the effective action. For instance, in the case (\ref{F30}) for the simplest descendant, we should assume that the second derivative of the analytic change of variables vanishes at the origin. 
\section{Modification of Ward identities}\label{Ward}
In this section we are interested in how the theory is modified if one nevertheless considers the functional integral over $h_{++}$ with the action (\ref{F22}). That is possible in the semiclassical approximation, which we use here. In the light-cone gauge the most efficient way to derive EOM and Ward identities is to go to the variables (\ref{F10}), because the substitution of (\ref{F10}) into (\ref{F22}) gives Polyakov-Liouville part of the effective action in local form:
\begin{equation}\label{F32}
	W_{\text{eff}}[f] = -\Delta \int d^2\boldsymbol{x} \delta^{(2)}(\boldsymbol{x}) \log\left(\partial_{-}f\right) - \frac{d}{24\pi}\int d^2\boldsymbol{x} \left[ \frac{\partial_{+}\partial_{-}f \partial_{-}^2 f}{\left(\partial_{-} f\right)^2} - \frac{\partial_{+}f \left(\partial_{-}^2 f\right)^2}{\left(\partial_{-} f\right)^3}\right].
\end{equation}
Under the gauge transformation (\ref{F11}) we find the variation of this action:
\begin{equation}\label{F33}
	\delta W_{\text{eff}} = \Delta \int d^2\boldsymbol{x}\left[ \partial_{-}\delta^{(2)}(\boldsymbol{x}) - \delta^{(2)}(\boldsymbol{x})\partial_{-}\log(\partial_{-} f) \right]\epsilon_{+}(x) + \frac{c}{12\pi}\int d^2\boldsymbol{x} \partial_{-}^3 h_{++}\epsilon_{+}(x),
\end{equation}
where $c = 26-d$ and reflects the determinant of the operator in r.h.s. of (\ref{F11}) \cite{Polyakov2d}. Now we go back to $h_{++}$ as the dynamical variable and write down modified Ward identities:
\begin{gather}
	\frac{c}{12\pi}\partial_{-}^3 \Big\langle h_{++}(z)h_{++}(x_1)...h_{++}(x_N)\Big\rangle + \Delta \partial_{-}\delta^{(2)}(\boldsymbol{z})\Big\langle h_{++}(x_1)...h_{++}(x_N)\Big\rangle - \nonumber \\
	- \Delta\delta^{(2)}(\boldsymbol{z})\Big\langle \bigg(\partial_{-}\log\left( \partial_{-}f(z)\right)\bigg)h_{++}(x_1)...h_{++}(x_N)\Big\rangle = \text{r.h.s. of (\ref{F13})}.
\label{F34}
\end{gather}
Note that (\ref{F34}) is an integro-differential equation unlike (\ref{F13}), because we should expand $f[h_{++}]$ in all orders in $h_{++}$. As it was already pointed out above, in the present paper we postpone the discussion of the influence of quantum fluctuations of $h_{++}$ and formally take the quasiclassical limit $d\rightarrow \infty,\; c\simeq -d$, because even under this assumption we can make some conclusions. Indeed, let us linearize the lowest order Ward identity (\ref{F34}), which is nothing but the averaged equation of motion. Substitution of the expansion
\begin{equation}\label{F35}
	f[h_{++}] (x) = x_{-} - \frac{1}{\pi}\int d^2\boldsymbol{y} \frac{1}{y_{-}-x_{-}}h_{++}(y) + \ldots
\end{equation}
into (\ref{F34}) gives
\begin{equation}\label{F36}
	\frac{d}{12\pi}\partial_{-}^3\langle h_{++}(x)\rangle \simeq \Delta \partial_{-}\delta^{(2)}(\boldsymbol{x}) + \frac{2\Delta}{\pi}\delta^{(2)}(\boldsymbol{x})\int d^2\boldsymbol{y} \frac{1}{y_{-}^3}\langle h_{++}(y)\rangle.
\end{equation}
In equations (\ref{F35})--(\ref{F36}) we omit higher order correlators in the expansion due to the fact that they are suppressed in the quasiclassical limit: $\langle h_{++}(x)h_{++}(y) \rangle \sim \frac{1}{d}$ etc. Analogously to the case of (\ref{F13}) we can use the relation in two dimensions
\begin{equation}\label{F37}
	\delta^{(2)}(\boldsymbol{x}) = \frac{1}{\pi} \partial_{-}^2 \frac{x_{-}}{x_{+}} = \frac{1}{2\pi} \partial_{-}^3 \frac{x_{-}^2}{x_{+}}
\end{equation} 
and rewrite (\ref{F36}) in the form
\begin{equation}\label{F38}
	\frac{d}{12\pi}\langle h_{++}(x)\rangle = \frac{\Delta}{\pi} \frac{x_{-}}{x_{+}} + \frac{\Delta}{\pi^2}\int d^2\boldsymbol{y} \frac{1}{y_{-}^3}\langle h_{++}(y)\rangle.
\end{equation}
Note that $h_{++} = 0$ is no more a solution of the equation of motion. Instead, we can set
\begin{equation}\label{F39}
	\langle h_{++}(x)\rangle = \frac{12\Delta}{d}\frac{x_{-}}{x_{+}},
\end{equation}
which solves (\ref{F38}) in view of the fact that
\begin{equation*}
   \int \frac{1}{y_{-}^3}\frac{y_{-}}{y_{+}}\frac{dy_{+}dy_{-}}{2i} = \int e^{-\tau-i\sigma}d\tau d\sigma = 0,  
\end{equation*}
where we've also introduced space-time coordinates $ x_{\pm} = e^{\tau\mp i\sigma}$. The situation looks more schematically if we move to Minkowski signature: $\sigma\rightarrow i\sigma$.  Then the averaged metric tensor $\langle g_{\mu\nu}\rangle$, given by the solution (\ref{F39}), induces the following geometry:
\begin{equation}\label{F40}
	ds^2 = e^{2\tau}\left[ \left(1+\frac{12\Delta}{d}\right)d\tau^2 - \left(1 -\frac{12\Delta}{d} \right)d\sigma^2 + \frac{24\Delta}{d}d\tau d\sigma \right].
\end{equation}
Its eigenvalues:
\begin{equation*}
    \lambda_{\pm} = \frac{12\Delta}{d} \pm \sqrt{1+\left(\frac{12\Delta}{d}\right)^2},
\end{equation*}
so the signature is unperturbed, and the space become flat in the limit $d\rightarrow\infty$ or $\Delta \rightarrow 0$, as it should be. Nevertheless, we can see that various states of conformal matter coupled to gravity in (\ref{F14}) can lead to some deformations of averaged space geometry predictably like (\ref{F40}). It seems quite curious which state insertions lead to which geometries, and how many deviations from flat space we may observe.  
\section{Thermal conformal matter}\label{Thermal}
In this section, we outline our approach in the situation of thermalized conformal matter. For definiteness, we work with bosonic field. In \cite{ThermalCFT} the thermal Wightman function for the massless conformal field in flat space was constructed:
\begin{equation}\label{F41}
	W_{\beta}(x,y) = \langle \phi(x)\phi(y)\rangle_{\beta} = \text{const} - \frac{1}{4\pi}\log\left( -\sinh\left(\frac{\pi(x_{-}-y_{-})}{\beta}\right)\sinh\left(\frac{\pi(x_{+}-y_{+})}{\beta}\right) \right),
\end{equation}
where
\begin{equation}
    \langle \mathcal{O} \rangle_{\beta} = \frac{\text{tr}\left\{\mathcal{O} e^{-\beta \hat{H}}\right\}}{\text{tr} \left\{e^{-\beta \hat{H}}\right\}}
\end{equation}
and $\beta$ is the inverse temperature. Note that the expression (\ref{F41}) is well-defined on flat Minkowski space with real-valued variables $\left( x_{+}, x_{-} \right)$, while we should restrict the domain to the tube 
\begin{equation}\label{F42}
 \left\{\left(x_{+}, x_{-}\right): \; -\beta < \text{Im}\left(x_{+}\right) < 0, \; -\beta < \text{Im}\left(x_{-}\right) < 0\right\}
\end{equation}
when we work in Euclidean signature (see also \cite{ThermalityInCft}). Then we consider the following action with the interaction term:
\begin{equation}\label{F43}
	S_{\text{CFT}}\left[ \phi, h_{++}\right] = \frac{1}{2\pi}\int d^2\boldsymbol{x} \partial_{+}\phi(x) \partial_{-}\phi(x) + \frac{1}{\pi}\int d^2\boldsymbol{x} \left(-\frac{1}{2}\left(\partial_{-}\phi(x)\right)^2\right)h_{++}(x),
\end{equation}
where we assume the integration over the domain (\ref{F42}). Now we will calculate the series (\ref{F15}) using Wick theorem (it works with thermal averages as well). Then an arbitrary summand up to some common factor takes the form
\begin{gather}
	\int d^2x_1 \ldots d^2x_N h_{++}(x_1) \ldots h_{++}(x_N)\big\langle \partial_{-}\phi(x_{i_1})\partial_{-}\phi(x_{j_1}) \big\rangle \ldots \big\langle \partial_{-}\phi(x_{i_N})\partial_{-}\phi(x_{j_N}) \big\rangle = \nonumber
	\\
	= \left(\frac{-1}{4\pi}\right)^N \left(\frac{\pi}{\beta}\right)^{2N} \int d^2 x_1 \ldots d^2 x_N h_{++}(x_1) \ldots h_{++}(x_N)\frac{1}{\sinh^2\left(\frac{\pi}{\beta}(x_{i_1}^{-}-x_{j_1}^{-})\right)} \ldots \frac{1}{\sinh^2\left(\frac{\pi}{\beta}(x_{i_N}^{-}-x_{j_N}^{-})\right)},
\label{F44}
\end{gather}
for some permutations $\{i_k\},$ and $\{j_k\}$ of $1,\ldots,N$ such that $i_k\neq j_k$. Now there is no surprise that by using ``cylinder to plane'' change of variables $\xi = e^{\frac{2\pi}{\beta}x}$ and the relation
\begin{equation}\label{F45}
	2\frac{1}{\xi_1-\xi_2} = e^{-\frac{\pi}{\beta}x_1}e^{-\frac{\pi}{\beta}x_2}\frac{1}{\sinh\left( \frac{\pi}{\beta}(x_1-x_2)\right)},
\end{equation}
one can find that (\ref{F44}) is equivalent to
\begin{equation}\label{F46}
	\int d^2\xi_1\ldots d^2\xi_N \tilde{h}_{++}(\xi_1)\ldots \tilde{h}_{++}(\xi_N) \frac{1}{(\xi_{i_1}-\xi_{j_1})^2}\ldots \frac{1}{(\xi_{i_N}-\xi_{j_N})^2},
\end{equation}
where we denote
\begin{equation}\label{F47}
	\tilde{h}_{++}\left(\xi(x)\right) = \frac{2\pi}{\beta} e^{\frac{2\pi}{\beta}x_{-}} h_{++}(x). 
\end{equation}
Therefore, from (\ref{F44}) we have obtained the corresponding term (\ref{F46}) of the expansion (\ref{F8}), formulated in terms of different coordinates. The answer for the whole series of terms of the form (\ref{F46}) sums up to (\ref{F8}) with
\begin{equation}\label{F48}
		W_{\text{eff}} = \frac{1}{24\pi}\int d^2\xi d^2\eta \partial_{-}^2 \tilde{h}_{++}(\xi)\frac{1}{\partial_{\xi_{-}}\left(\partial_{\xi_{+}}-\tilde{h}_{++}\partial_{\xi_{-}}\right)}\partial_{\eta_{-}}^2 \tilde{h}_{++}(\eta).
\end{equation}
Going back to the light-cone coordinates, one can find that
\begin{equation}\label{F49}
	W_{\text{eff}} = \frac{1}{24\pi}\int d^2\boldsymbol{x}d^2\boldsymbol{y} \left(\partial_{-}^2 h_{++}(x) + \frac{2\pi}{\beta}\partial_{-}h_{++}(x)\right)\left[\frac{1}{\partial_{-}\left(\partial_{+}-h_{++}\partial_{-}\right)}\right]_{\beta}\left(\partial_{-}^2 h_{++}(y) + \frac{2\pi}{\beta}\partial_{-}h_{++}(y)\right),
\end{equation}
where index $\beta$ stands for appropriate periodic conditions on the laplacian in the tube (\ref{F42}). There are strange terms with the first derivative of $h_{++}$ appear in the parentheses in (\ref{F49}). In fact, their appearance is the result of the change of variables -- the metric transformation reads:
\begin{equation}\label{F50}
	||g_{\mu\nu}(x)|| = \begin{pmatrix}
	h_{++} & \frac{1}{2}\\
	\frac{1}{2} & 0
	\end{pmatrix} \rightarrow ||g_{\mu\nu}\left(\xi(x)\right)|| = \frac{2\pi}{\beta}e^{\frac{2\pi}{\beta}x_{-}}\begin{pmatrix}
	\frac{2\pi}{\beta}e^{\frac{2\pi}{\beta}x_{-}}h_{++} & \frac{1}{2}\\
	\frac{1}{2} & 0
	\end{pmatrix}.
\end{equation}
Then, the common scale factor in (\ref{F50}) shifts Ricci scalar:
\begin{equation}\label{F51}
	\sqrt{|\tilde{g}|}\tilde{R} = \sqrt{|g|}\left[R - \Box\left( \frac{2\pi}{\beta}x_{-} + \log\frac{2\pi}{\beta} \right)\right] = 2\left[\partial_{-}^2 h_{++} + \frac{2\pi}{\beta}\partial_{-}h_{++}\right].
\end{equation}
These terms do not vanish in (\ref{F49}), because the analytic changes are not included in the residual symmetry of the light cone gauge. The analysis above shows that the situation with conformal matter immersed into the thermal state is not really of big difference from the case with trivial density matrix. Nevertheless, the answer (\ref{F49}) provides us with the perspective to investigate the situation at the quantum level of $h_{++}$.
\section{Conclusion}\label{Concl}
Our main goal was to show that the effects of non-trivial initial states lead to new considerable phenomena on the landfill of induced two-dimensional quantum gravity, introduced by A.M.Polyakov in classical works \cite{Polyakov2d, KPZ, BosonStr}. First, we considered the theory induced by conformal matter with averages over spinless primary state, which is defined as the insertion of the operator at the origin:
$$
|\Delta\rangle \overset{\text{def}}{=} \lim\limits_{z\rightarrow 0} \Phi_{\Delta}(z)|0\rangle, \;\;z=e^{\tau-i\sigma},
$$
where $\tau$ has an interpretation of time in Hamiltonian approach in CFT. In general, perturbation at the origin may include the whole set of states of conformal matter, but these effects are more complicated (see section \ref{Descendants}) and we leave them for further work. However, it doesn't affect the general conclusion, that the gravitational background $\langle g_{\mu\nu}(x)\rangle$ gets a non-trivial behavior with some source, placed at past infinity $\tau\rightarrow -\infty$ as we see from (\ref{F36}), (\ref{F39})--(\ref{F40}). This behavior can be estimated with the use of Ward identities (\ref{F34}). Next, we introduce the effective action for gravity, induced by conformal matter immersed in thermal state with an appropriate Wightman function (\ref{F41}). It's worth noting that these two examples are just the simplest choices of Poincare non-invariant states one can consider. There are many other states that someone might be interested in, e.g. inclusions of conformal operators at different points of the manifold $M$ into the functional integral over conformal fields (\ref{F1}) or density matrices, constructed as distributions in momentum space etc. 

Finally, let us motivate the reason why we use $h_{++}$ as the fluctuating variable with linear functional measure. Despite the fact that the work of Distler and Kawai provided us with a powerful conjecture \cite{Distler:1988jt} to formulate Polyakov-Liouville theory in terms of functional integral with linear measure for the Liouville field, the original papers \cite{Polyakov2d, KPZ, LesHouches} introduce a dynamic mechanism to work with $2d$ gravity at quantum level. Namely, what we have done in this paper is that we've found the sum of the series (\ref{F14})--(\ref{F15}) with averages over conformal fields, interacting with some external background field $\hat{h}_{++}$. In general, we should proceed and study the corrections of quantum fluctuations: $h_{++}(x) = \hat{h}_{++}(x) + a(x)$. Then in one loop order one obtains that
\begin{equation*}
	\mathcal{Z}_{\text{full}} = \int \mathscr{D}h_{++} e^{-W_{\text{eff}}[h_{++}]} \simeq e^{-W_{\text{eff}}[\hat{h}_{++}]}\int \mathscr{D}a\; \text{exp}\left\{-a\left[\frac{\delta^2}{\delta a \delta a} W_{\text{eff}}\right] a\right\}
\end{equation*} 
and gets in the original approach the dynamical reason for the renormalization of $k_{0}$ in (\ref{F9}). However, we can have more constants which appear in density matrix definition, such that in order to complete our considerations we need to carefully conduct the treatment of the theory at the quantum level for different $\hat{\rho}_{0}$ in (\ref{F14}) and find possible modifications, if any. Finally, we hope that our work will lead to a slightly wider view of the Universe in two dimensions, which can be thought of as a toy model for exploring our methods in different situations and theories.

\section*{Acknowledgements}
We would like to thank I.Kochergin, A.Artemev for careful reading of the paper and useful discussions. Especially we are grateful to E.T.Akhmedov for initiating this work, sharing ideas and support. This work was supported by the grant from the Foundation for the Advancement of Theoretical Physics and Mathematics ``BASIS'' and by Russian Ministry of education and science.

\appendix
\section{Green function}\label{AppA}
In this  appendix we show how to compute the Green function $\frac{1}{\Box}(x,y)$ (\ref{F6}) perturbatively in $h_{++}$. First, rewrite the definition (\ref{F6}) in the light-cone gauge:
\begin{equation}\label{A1}
    4\partial_{-}\left( \partial_{+} - h_{++}\partial_{-}\right)\frac{1}{\Box}(x,y) = \delta^{(2)}(\boldsymbol{x}-\boldsymbol{y}).
\end{equation}
At zeroth order:
\begin{equation}\label{A2}
    4\partial_{-}\partial_{+}\frac{1}{\Box}^{(0)}(x,y) = \delta^{(2)}(\boldsymbol{x}-\boldsymbol{y}), \quad \text{hence} \quad
    \frac{1}{\Box}^{(0)}(x,y) = \frac{1}{4\pi} \log\left( (x_{+}-y_{+})(x_{-}-y_{-}) \right),
\end{equation}
where we use the property
\begin{equation}\label{A3}
    \partial_{-}\frac{1}{x_{+}} = \partial_{+}\frac{1}{x_{-}} = \pi \delta^{(2)}(\boldsymbol{x}).
\end{equation}
In order to determine the first correction we substitute the decomposition $\frac{1}{\Box} = \frac{1}{\Box}^{(0)} +  \frac{1}{\Box}^{(1)}$ into (\ref{A2}) and obtain:
\begin{equation}\label{A4}
    4\partial_{-}\partial_{+}\frac{1}{\Box}^{(1)}(z,y) =  -4\partial_{-}\left( h_{++}\partial_{-}\right)\frac{1}{\Box}^{(0)}(z,y)
\end{equation}
Then one multiplies both sides by $\frac{1}{\Box}^{(0)}(x,z)$, integrates the equation by parts over $z$ to get
\begin{equation}\label{A5}
    \frac{1}{\Box}^{(1)}(x,y) = \frac{1}{4\pi^2}\int d^2z \frac{1}{x_{-}-z_{-}}\frac{1}{z_{-}-y_{-}} h_{++}(z).
\end{equation}
Similarly for the second order:
\begin{equation}\label{A6}
    \frac{1}{\Box}^{(2)}(x,y) = -\frac{1}{4\pi^3}\int d^2zd^2\zeta \frac{1}{x_{-}-z_{-}} \frac{1}{(z_{-}-\zeta_{-})^2} \frac{1}{z_{-}-y_{-}} h_{++}(z)h_{++}(\zeta).
\end{equation}
This procedure can be easily continued and one can write down the general expression for any order in $h_{++}$ by induction.

Also, we can write a closed expression for the Green function by the use of the variable $f$, which is defined as $\partial_{+}f - h_{++}\partial_{-}f = 0$, such that
\begin{equation}\label{A7}
    \partial_{-}\left( \partial_{+} - h_{++}\partial_{-}\right) \log \left(x_{+}f(x) \right) = \pi \delta^{(2)}(\boldsymbol{x}) + \partial_{-}\left[\frac{\partial_{+}f - h_{++}\partial_{-}f}{f} \right] = \pi \delta^{(2)}(\boldsymbol{x}).
\end{equation}
Hence, the closed expression is given by 
\begin{equation}\label{A8}
	\frac{1}{\Box}(x,y) = \frac{1}{4\pi}\log\big\{(x_{+} - y_{+})\left(f(x)-f(y)\right)\big\}.
\end{equation}

\printbibliography

\end{document}